\documentclass[aps,prb,twocolumn,superscriptaddress,groupedaddress]{revtex4-1} 

\usepackage[dvipdfm]{graphicx} 
\usepackage{dcolumn} 
\usepackage{amsmath,txfonts,bm}

\usepackage{color}
\usepackage{ulem}   
\usepackage{comment}

\begin{document}

\widetext

\title{ Possibility of n-type doping in CaAl$_2$Si$_2$-type Zintl phase compound CaZn$_2X_2$ ($X$ = As,  P) } 


\author{ 
Kazutaka Nishiguchi,$^1$ 
Masayuki Ochi,$^1$ 
Chul-Ho Lee,$^2$ 
and Kazuhiko Kuroki$^1$ 
} 
\affiliation{ 
$^1$Department of Physics, Osaka University, 1-1 Machikaneyama-cho, Toyonaka, Osaka 560-0043, Japan  \\ 
$^2$National Institute of Advanced Industrial Science and Technology (AIST), Tsukuba, Ibaraki 305-8568, Japan
} 
                             
\date{\today}

\begin{abstract} 
Motivated by a recent theoretical suggestion that doping electrons into various CaAl$_2$Si$_2$-type Zintl phase compounds may give rise to high thermoelectric performance, 
we explore the possibility of n-type (electron carrier) doping of CaZn$_2X_2$ ($X$ = As, P) using first principles calculation. 
We consider n-type doping of CaZn$_2X_2$ with the following two situations: 
interstitial-site doping of alkaline earth metals $AE$ (= Mg, Ca, Sr, Ba) and group 3 elements $G3$ (= Sc, Y, La), and $G3$ substitutional doping for the Ca site. 
The evaluation of the formation energy of these defects in various charged states reveals that 
the interstitial-site doping of $AE$ = Ca/Mg or $G3$ = Sc/Y, and $G3$ = La/Y substitutional doping for the Ca site are relatively favorable among the possibilities considered. 
In particular, the formation energy of the La substitutional doping for the Ca site is the lowest among the considered cases both for CaZn$_2X_2$ ($X$ = As, P) 
and is negative, which implies that La is expected to be substituted for the Ca site and provide electron carriers spontaneously. 
We also find that for each doping case considered, the formation energy of the defects is smaller for $X$ = As than for $X$ = P, 
which suggests that former is relatively easier to realize n-type doping than the latter. 
\end{abstract}

\maketitle

\section{Introduction} 
Thermoelectric devices enable us to directly convert waste heat into electric current, 
so that not only thermoelectric (Seebeck) effect is one of the most fundamental and intriguing quantum mechanical phenomena in condensed matter physics, 
but also thermoelectric materials are important applications in terms of clean and sustainable energy.~\cite{ZHANG201592,Kauzlarich_2007} 
Conversion efficiency of thermoelectric generators is evaluated by the dimensionless figure of merit $ZT = \sigma S^2 T / \kappa$ and power factor $PF = \sigma S^2$, 
where $\sigma$ is the electrical conductivity, $S$ is the Seebeck coefficient, $\kappa$ is the thermal conductivity, and $T$ is the absolute temperature. 
The thermal conductivity can be usually divided into the lattice and carrier contributions ($\kappa = \kappa_{\text{lat}} + \kappa_{\text{el}}$), 
so that the phonon structures as well as the electronic structures play an important role for thermoelectric performance. 

A simple strategy to achieve high thermoelectric efficiency is to realize both of large power factor and small thermal conductivity at the same time. 
Nanostructuring can suppress the lattice thermal conductivity $\kappa_{\text{lat}}$ while keeping high carrier mobility.~\cite{Hochbaum_2008,Boukai_2008,Poudel_2008} 
On the other hand, it is in general difficult to enhance $PF$ 
because $\sigma$ and $S$ are difficult to increase simultaneously 
due to the trade-off relation between them against the carrier concentration. 

In terms of electronic band structures, 
high group velocity of electrons and large density of states (DOS) around the Fermi level enable us to enhance $PF$. 
There have been several theoretical proposals for specific band structures that can give rise to large power factor: 
multi-valley~\cite{Pei_2011}, pudding-mold-type~\cite{Kuroki_2007}, and low dimensional band structures.~\cite{Hicks:1993tj,Hicks:1993tq,Dresselhaus_2007} 
Especially, multi-valley band structures are suitable for designing high performance thermoelectric materials, 
where chemical doping and substitution enable us to control the valley degeneracy through the change in the electronic and crystal structure. 
In fact, high performance thermoelectric materials $\alpha$-MgAgSb ($ZT = 1.2$--$1.4$ at 550 K)~\cite{Liu_2018} and Mg$_{3+\delta}$(Sb,Bi,Te)$_2$ ($ZT = 1.5$ at 716 K)~\cite{Tamaki_2016,Ohno_2018,Shi_2019} are known to be good examples that have multi-valley band structures. 

In recent years, Zintl phase compounds have attracted much interest both experimentally and theoretically as high performance thermoelectric materials.~\cite{SHUAI201774,Brown_2006,Ohno_2017,Wang_2009,Tamaki_2016} 
Among them, Yb(Cd,Zn)$_{2}$Sb$_2$~\cite{Wang_2009} and Mg$_{3+\delta}$(Sb,Bi,Te)$_2$~\cite{Tamaki_2016,Ohno_2018,Shi_2019} have large $ZT > 1$, 
and they are 122 Zintl phase compounds $AB_2X_2$, whose $A$ is alkaline earth metal or lanthanoid, $B$ is Mg, Zn, or Cd, and $X$ is pnictogen. 
122 Zintl phase compounds have CaAl$_2$Si$_2$-, BaCu$_2$S$_2$-, or ThCr$_2$Si$_2$-type crystal structure,~\cite{Peng_2018} 
and especially the CaZn$_2$Si$_2$-type structure with the $P\bar{3}m1$ space group often exhibits large $ZT > 1$ in both the n-type and p-type compounds.  
It is understood that high thermoelectric performance of the CaZn$_2$Si$_2$-type compounds can be caused by the multi-valley band structure.~\cite{Zhang_2017,Zhang_2019,Toberer_2010} 

Very recently, thermoelectric performance of many kinds of CaAl$_2$Si$_2$-type Zintl phase compounds has been theoretically estimated by first-principles band calculations.~\cite{Usui_2020} 
This work suggests that CaZn$_2X_2$ ($X$ = Sb, As, P) gives large $ZT$ in the electron-doped regime due to its high valley degeneracy of the electronic band structure. 
Indeed, this kind of the 122 Zintl phase compounds have been synthesized~\cite{Gascoin_2005,May_2012,Kihou_2017} 
and their electronic structure have also been investigated theoretically.~\cite{Toberer_2010,Sun_2017}
However, these 122 Zintl phase compounds become p-type semiconductors 
because the cation-site vacancies are easily formed under usual synthesis conditions.~\cite{Tamaki_2016,Ohno_2018,Pomrehn_2014,Peng_2018}  
Considering that high-performance thermoelectric material Mg$_{3+\delta}$(Sb,Bi,Te)$_2$ was successfully realized by n-type doping,~\cite{Tamaki_2016,Ohno_2018,Shi_2019} 
it is worth investigating whether it is possible to introduce electron carriers into CaZn$_2X_2$. 

Motivated by the above background, we explore the possibility of n-type doping of CaAl$_2$Si$_2$-type Zintl phase compounds CaZn$_2X_2$ ($X$ = As, P) 
using first-principle calculations based on the density functional theory (DFT). 
The possibility of n-type (electron carrier) doping into CaZn$_2X_2$ is considered with the following two situations: 
interstitial-site doping of the alkaline earth metals $AE$ (= Mg, Ca, Sr, Ba) and group 3 elements $G3$ (= Sc, Y, La), and $G3$ substitutional doping for the Ca site. 
To see this, the formation energy of the chemical doping is evaluated numerically with the supercell approach.
We can find that the interstitial-site doping of $AE$ = Ca, Mg or $G3$ = Sc, Y, 
and $G3$ = La, Y substitutional doping for the Ca site is favorable both for CaZn$_2X_2$ ($X$ = As, P) in terms of energy stability. 
In particular, the formation energy of the La substitutional doping for the Ca site is the lowest among the considered cases both for CaZn$_2X_2$ ($X$ = As, P) 
and is negative so that La is expected to be substituted for the Ca site and provide electron carriers spontaneously. 
We can also see that each formation energy for CaZn$_2$As$_2$ is smaller than that for CaZn$_2$P$_2$ as a whole, 
so that n-type doping is considered to be relatively easier in CaZn$_2$As$_2$ than in CaZn$_2$P$_2$.

\section{Method}

\subsection{Theory and Model}  
In order to describe dopants and defects in a crystal under periodic boundary conditions, the supercell approach is adopted in the DFT calculations: 
we numerically investigate $N_{1} \times N_{2} \times N_{3}$ ($N_{1,2,3} \in \mathbb{N}$) supercells 
whose supercell lattice vectors are $\bm{A}_{i} = N_{i} \bm{a}_{i}$ ($i=1,2,3$) with $\bm{a}_{i}$ being the $i$-th primitive lattice vector. 
In this study, the atomic positions in the supercell are relaxed in the structural optimization 
after the lattice constants and atomic positions of the primitive cell are optimized. 
To consider the possibility of chemical doping, 
the dopant/defect formation energy of charged states is calculated using the supercells. 
The formation energy can be obtained from the supercell calculations,~\cite{Zhang_1991,Komsa_2012} as 
\begin{equation} 
E_{\text{form}} [D^q] = \{ E[D^q] +E^{\text{corr}}[D^q] \} -E_{\text{P}} -\sum_{i} n_{i} \mu_{i} +qE_{\text{F}}  \, ,  \label{eq:Eform} 
\end{equation} 
where $E[D^q]$ and $E_{\text{P}}$ denote the total energy of the supercell with dopant $D^q$ in charged state $q$ 
and that of the perfect crystal supercell without any dopants and defects, respectively. 
$E^{\text{corr}}[D^q]$ represents its image-charge correction as discussed below. 
Also, $n_{i}$ represents the number of removed or added atom $i$ with the chemical potential $\mu_{i}$, and $E_{\text{F}}$ is the Fermi level, i.e., the chemical potential of the charge reservoir. 

Since the periodicity of systems under periodic boundary conditions is an inevitable artifact in the supercell approach, 
we evaluate the formation energy in the dilute limit by including the energy correction of the finite size effects. 
For a cubic (isotropic) supercell, the image-charge correction in charged state $q$, $E^{\text{corr}}_{\text{iso}}$, can be written as a function of the linear dimension of the supercell $L$, 
\begin{equation} 
E^{\text{corr}}_{\text{iso}} = \frac{q^2 \alpha}{2\varepsilon L} -\frac{2\pi qQ}{3\varepsilon L^3} +\frac{2\pi \bm{p}^{2}}{3\varepsilon L^3} +O(L^{-5}). 
\label{eq:finite_size} 
\end{equation} 
Furthermore, $\alpha$, $\varepsilon$, $\bm{p}$, and $Q$ represents the Madelung constant, dielectric constant, and dipole and quadrupole moment of the system.~\cite{Leslie_1985,Makov_1995,Fraser_1996,Dabo_2008} 
Note that here the dielectric constant $\varepsilon$ is a scalar quantity since the expression in Eq. (\ref{eq:finite_size}) is the result for cubic (isotropic) systems. 
This quantity should be treated as a dielectric tensor $\hat{\varepsilon}$ when one considers anisotropic systems like the 122 Zintl phase compound CaZn$_2X_2$ ($X$ = As, P) in this study. 
In particular, the counterpart of the second term in Eq. (\ref{eq:finite_size}), namely, the point-charge correction $q^2 \alpha / 2\varepsilon L$, for anisotropic systems can be given as~\cite{Ziman_1972,Rurali_2009,Kumagai_2014} 
\begin{equation} 
\begin{split} 
& E^{\text{corr}}_{1/L} 
= -\frac{q^2}{2} \Bigg[ 
\sum_{\bm{R}_{i} \neq \bm{0}} \frac{1}{\sqrt{\mathrm{det} \, \hat{\varepsilon}}} \frac{\mathrm{erfc}(\eta \sqrt{\bm{R}_{i} \cdot \hat{\varepsilon}^{-1} \cdot \bm{R}_{i}})}{\sqrt{\bm{R}_{i} \cdot \hat{\varepsilon}^{-1} \cdot \bm{R}_{i}}}  \\ 
&\qquad \qquad \quad 
+\frac{4\pi}{\Omega} \sum_{\bm{G}_{i} \neq \bm{0}} \frac{\mathrm{exp} \left( -\bm{G}_{i} \cdot \hat{\varepsilon} \cdot \bm{G}_{i} /4\eta^2 \right)}{\bm{G}_{i} \cdot \hat{\varepsilon} \cdot \bm{G}_{i}}  \\ 
&\qquad \qquad \quad 
-\frac{2\eta}{\sqrt{\pi \, \mathrm{det} \, \hat{\varepsilon}}} -\frac{\pi}{\eta^2 \Omega} 
\Bigg] ,  \label{eq:1/L-term} 
\end{split} 
\end{equation} 
where the sum over $\bm{R}_{i}$ ($\bm{G}_{i}$) extends over all vectors of the supercell direct (reciprocal) lattice except for $\bm{R}_{i} (\bm{G}_{i}) = \bm{0}$, 
$\Omega = |\bm{A}_{1} \cdot \left( \bm{A}_{2} \times \bm{A}_{3} \right)| = N_{1}N_{2}N_{3} V$ denotes the supercell volume 
with $V = |\bm{a}_{1} \cdot \left( \bm{a}_{2} \times \bm{a}_{3} \right)|$ being the primitive cell volume, 
and $\eta$ is a suitably chosen convergence factor ($\sim \mathrm{min} \{ |\bm{G}_{i}| \}$). 

In this study, 
the formation energy in the dilute limit is evaluated by extrapolating the numerical results of the supercell calculations as a function of the linear dimension of the supercell $L = \sqrt[3]{\Omega}$, 
where the point-charge correction $E^{\text{corr}}_{1/L}$ is taken into account. 
$E^{\text{corr}}_{1/L}$ can be obtained from the macroscopic static dielectric tensor $\hat{\varepsilon}$ through Eq. (\ref{eq:1/L-term}). 
The macroscopic static dielectric tensor including the ionic contribution is calculated based on the density functional perturbation theory, 
where the local field effect with respect to the Hartree and the exchange-correlation potential is included. 
The derivative of the cell-periodic part of the Kohn-Sham orbitals are calculated using the finite-difference method. 

\subsection{Numerical Methods} 
To numerically evaluate the formation energy of chemical doping in CaZn$_2X_2$, 
first-principles calculations based on DFT are performed within the generalized gradient approximation (GGA). 
We perform spin-polarized DFT calculations with the Perdew--Burke--Ernzerhof (PBE) exchange-correlation functional~\cite{PBE_1996} and the projector augmented wave (PAW) method.~\cite{Blochl_1994,Kresse_1999} 
In the PAW potentials, 
core-electron orbitals are set for each element as Ca:[Ne], Zn:[Ar], As:[Ar]3$d^{10}$, P:[Ne], Mg:[He], Sr:[Ar]3$d^{10}$, Ba:[Kr]4$d^{10}$, Sc:[Ne], Y:[Ar]3$d^{10}$, and La:[Kr]4$d^{10}$, 
and the number of valence electrons $N_{\text{e}}$ is 10 for the alkaline earth metals $AE$, $N_{\text{e}} = 12$ for Zn, $N_{\text{e}} = 5$ for $X$ = As and P, and $N_{\text{e}} = 11$ for the group 3 elements $G3$. 
In this study, we use the Vienna Ab initio Simulation Package (VASP).~\cite{Kresse_1993,Kresse_1994,Kresse_1996a,Kresse_1996b} 
Here we set the plane-wave cutoff energy to be 500 eV, and the crystal structure is optimized until the residual Hellmann-Feynman forces become smaller than 0.01 eV/\AA, 
and the spin-orbit coupling is not taken into account in this study. 

In this study, the primitive cell is determined first by optimizing both of the lattice constants and atomic positions, 
where the Monkhorst-Pack (MP) 13 $\times$ 13 $\times$ 7 $k$-point grids in the Brillouin zone are taken in the structural optimization. 
In the supercell calculations, the $N_{1} \times N_{2} \times N_{3}$ supercell is constructed by setting the supercell lattice vectors $\bm{A}_{i} = N_{i} \bm{a}_{i}$ ($i=1,2,3$) with $\bm{a}_{i}$ being the obtained $i$-th primitive lattice vector, 
and then only the atomic positions are relaxed in the structural optimization. 
To obtain the formation energies in the dilute limit, 
the 3 $\times$ 3 $\times$ 2, 3 $\times$ 3 $\times$ 3, 4 $\times$ 4 $\times$ 2, and 4 $\times$ 4 $\times$ 3 supercells are considered for extrapolation, 
where gamma-centered (G) 6 $\times$ 6 $\times$ 3, MP 5 $\times$ 5 $\times$ 3, MP 5 $\times$ 5 $\times$ 3, and G 4 $\times$ 4 $\times$ 3 $k$-point grids are used respectively. 

When each of the formation energy of the charged states in the dilute limit is evaluated from the supercell calculations, 
the total energies of the supercell with dopant $D^q$ in charged state $q$ including the point-charge correction, $\{E[D^q] +E^{\text{corr}}_{1/L}\}$, 
are fitted by the least squares method as a function $E_{\infty} +C/L^3$. 
Here $E_{\infty}$ represents the desirable total energy in the dilute limit ($L \rightarrow \infty$), 
and $C$ is a fitting parameter arising from the higher-order correction of $O(1/L^3)$.

\section{Results} 
\subsection{Crystal and electronic structures of CaZn$_2X_2$ ($X$ = As, P)} 
\begin{figure}[h] 
\begin{tabular}{cc}
\begin{minipage}{.20\textwidth} 
\centering
\includegraphics[width=4.0cm,clip]{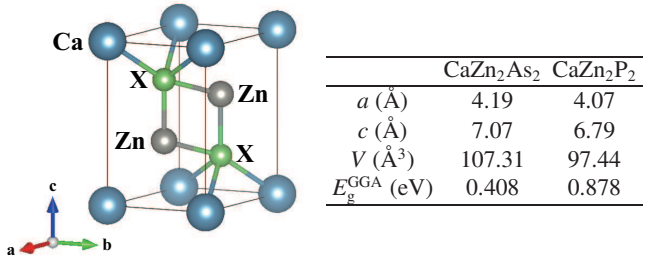} 
\end{minipage} 
\hfill
\begin{minipage}{.30\textwidth} 
\begin{center} 
\begin{tabular}{cccc} \hline 
                 & CaZn$_2$As$_2$  & CaZn$_2$P$_2$  \\ \hline
$a$ (\AA)  &  4.19  &  4.07  \\ 
$c$ (\AA)  &  7.07  &  6.79  \\ 
$V$ (\AA$^3$)  &  107.31  &  97.44  \\ 
$E^{\text{GGA}}_{\text{g}}$ (eV)  &  0.408  &  0.878  \\ \hline 
\end{tabular} 
\end{center} 
\end{minipage} 
\end{tabular} 
\caption{(Color online) 
(Left) Crystal structure of CaAl$_2$Si$_2$-type Zintl phase compound CaZn$_2X_2$ ($X$ = As, P). 
(Right) Their lattice constants $a$ and $c$, volume of the unit cell $V$, and band gap $E^{\text{GGA}}_{\text{g}}$ evaluated using GGA. 
} 
\label{fig:crystal_electronic}
\end{figure}

Before considering chemical doping, 
we first look at the crystal and electronic structures of CaZn$_2X_2$ ($X$ = As, P). 
The crystal structure, and the lattice constants $a$ and $c$, volume of the primitive cell $V$, and energy band gap $E^{\text{GGA}}_{\text{g}}$ evaluated using GGA are shown in FIG. \ref{fig:crystal_electronic}. 
All of the crystal structures shown in this paper are depicted by using the VESTA software.~\cite{Momma_2011} 
Each of the evaluated energy band gaps is the direct band gap at the $\Gamma$ point. 

Here the obtained band gap is underestimated since GGA usually provides a smaller band gap for semiconductors than the experimental one. 
In fact, although the experimental band gap for CaZn$_2$Sb$_2$ is finite ($\sim 0.26$ eV),~\cite{Toberer_2010} 
the evaluated band gap disappears within GGA ($E^{\text{GGA}}_{\text{g}}=0$). 
From now on, we do not refer to CaZn$_2$Sb$_2$ in our numerical calculations. 

In the same calculation conditions, 
the macroscopic static dielectric tensor $\hat{\varepsilon}$ is also calculated for obtaining the point-charge correction $E^{\text{corr}}_{1/L}$. 
Here $\hat{\varepsilon}$ is obtained from the summation over the electronic and ionic contribution, as 
\begin{equation} 
\hat{\varepsilon} = 
\begin{pmatrix} 
33.80 &   0.00 & 0.00  \\ 
 0.00  & 33.81 & 0.00  \\ 
 0.00  &   0.00 & 21.03
\end{pmatrix} 
\end{equation} 
for CaZn$_2$As$_2$, and 
\begin{equation} 
\hat{\varepsilon} = 
\begin{pmatrix} 
15.00 &   0.00 & 0.00  \\ 
 0.00  & 15.00 & 0.00  \\ 
 0.00  &   0.00 & 20.91
\end{pmatrix} 
\end{equation} 
for CaZn$_2$P$_2$.

\subsection{Interstitial-site doping of $AE$ and $G3$} 
We first consider the interstitial-site doping and their formation energies. 
Here the alkaline earth metals $AE$ (= Mg, Ca, Sr, Ba) and group 3 elements $G3$ (= Sc, Y, La) are chosen to be a dopant 
since they are expected to be cations $AE^{2+}$ and $G3^{3+}$, and hence expected to achieve n-type doping. 
This is an idea analogous to the one adopted for Mg$_{3+\delta}$(Sb,Bi,Te)$_2$, 
where excess Mg is introduced to the interstitial site of Mg$_3$Sb$_2$ so as to realize n-type doping.~\cite{Tamaki_2016,Ohno_2018} 

\begin{figure}[t]
\begin{center}
\includegraphics[width=8.5cm,clip]{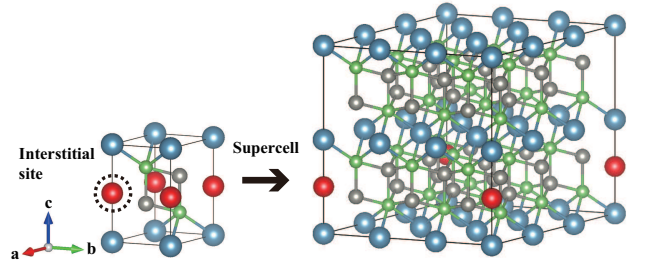} 
\caption{(Color online) 
(Left) Interstitial site of CaZn$_2X_2$ ($X$ = As, P). 
(Right) 3 $\times$ 3 $\times$ 2 supercell with a dopant (as an example). 
The red ball denotes the dopant, namely, $AE$ (= Mg, Ca, Sr, Ba) or $G3$ (= Sc, Y, La). 
} 
\label{fig:Supercell_interstitial}
\end{center} 
\end{figure}

The interstitial sites of CaZn$_2X_2$ and a supercell with a dopant (3 $\times$ 3 $\times$ 2 as an example) are shown in FIG. \ref{fig:Supercell_interstitial}. 
Since the formation energy of chemical doping is generally given by Eq. (\ref{eq:Eform}), 
we numerically evaluate the formation energy of the interstitial-site doping with dopant $D$ (= $AE$, $G3$) in this situation, 
\begin{equation} 
\begin{split} 
E_{\text{form}} 
&=         \{ E[D\text{:Ca$_N$Zn$_{2N}X_{2N}$}-qe^{-}] +E^{\text{corr}}_{1/L} \}  \\ 
&\quad -E[\text{Ca$_N$Zn$_{2N}X_{2N}$}] - E[D]  \\ 
&\quad +q\left( E_{\text{V}} + \Delta E_{\text{F}} \right) , 
\end{split} 
\end{equation} 
where $E[D\text{:Ca$_N$Zn$_{2N}X_{2N}$}-qe^{-}]$ and $E[\text{Ca$_N$Zn$_{2N}X_{2N}$}]$ 
are the total energy of the supercell of CaZn$_2X_2$ with dopant $D$ in a charged state with charge $q$ 
and that of the perfect crystal $N_1 \times N_2 \times N_3$ ($=N$) supercell without any dopants, respectively. 
In this study, the total energy (chemical potential) of dopant $D$, $E[D]$, is determined to be that of the bulk crystal per atom. 

\begin{figure}[t]
\begin{center}
\includegraphics[width=8.5cm,clip]{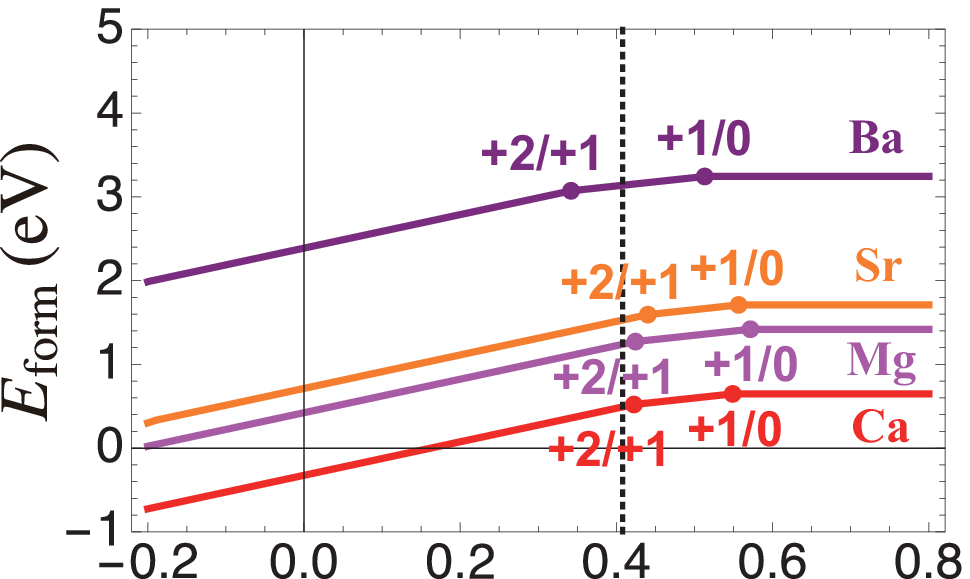} 
\includegraphics[width=8.5cm,clip]{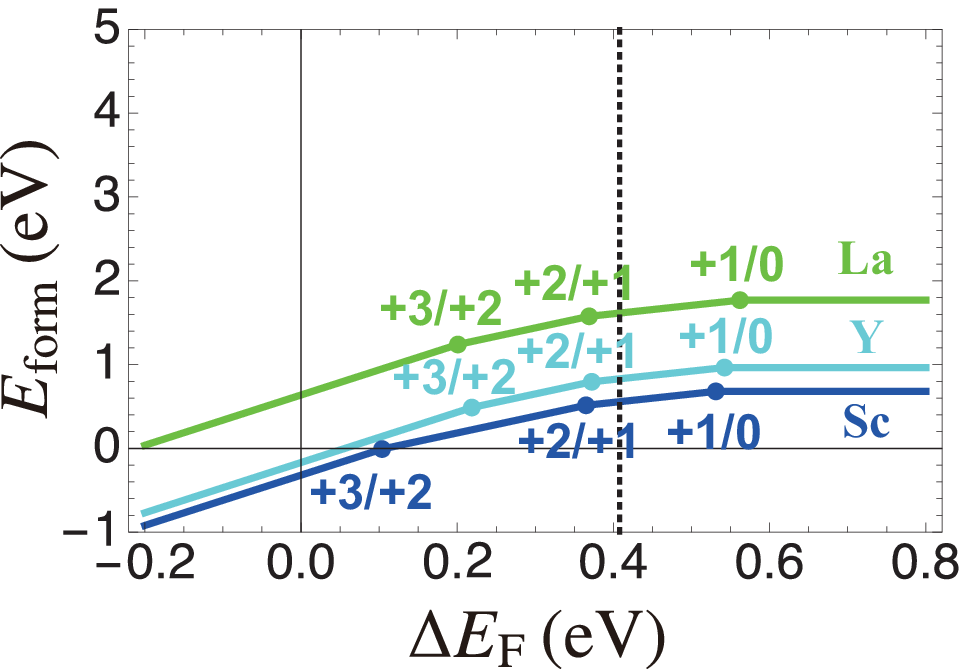} 
\caption{(Color online) 
The formation energy of the interstitial-site doping $E_{\text{form}}$ against the Fermi energy $\Delta E_{\text{F}} = E_{\text{F}} -E_{\text{V}}$ for CaZn$_2$As$_2$. 
(Top) The alkaline earth metal $AE$ (= Mg, Ca, Sr, Ba) and (Bottom) group 3 elements $G3$ (= Sc, Y, La) are chosen as dopant $D$. 
The least formation energy among different charged states ($q = 0, +1, +2, +3$) is plotted at each $\Delta E_{\text{F}}$, 
where the change of the charge is indicated at each points where the slope of the line changes. 
The dotted line denotes the position of the conduction band minimum ($\Delta E_{\text{F}} = 0.408$ eV). 
} 
\label{fig:xAs}
\end{center} 
\end{figure}

\begin{figure}[t]
\begin{center}
\includegraphics[width=8.5cm,clip]{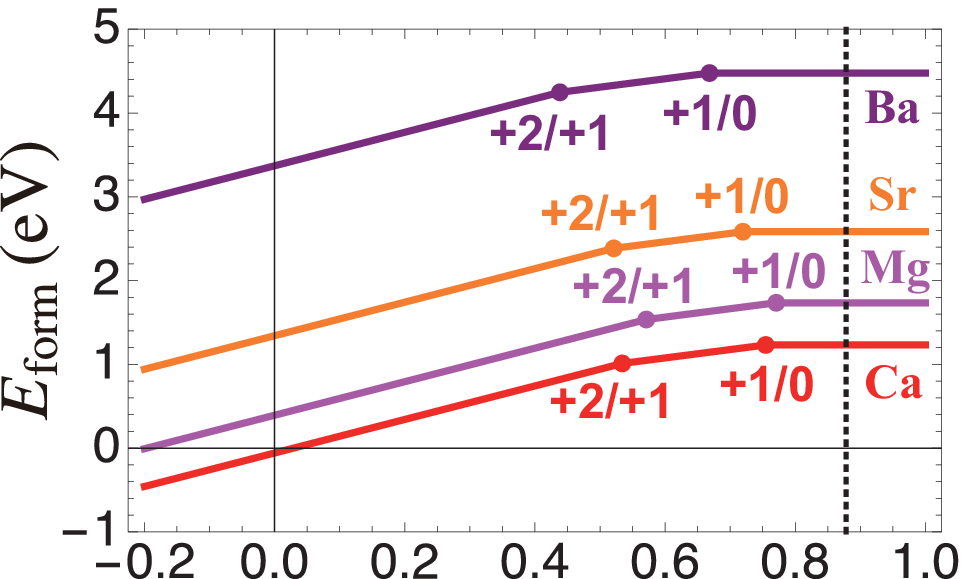} 
\includegraphics[width=8.5cm,clip]{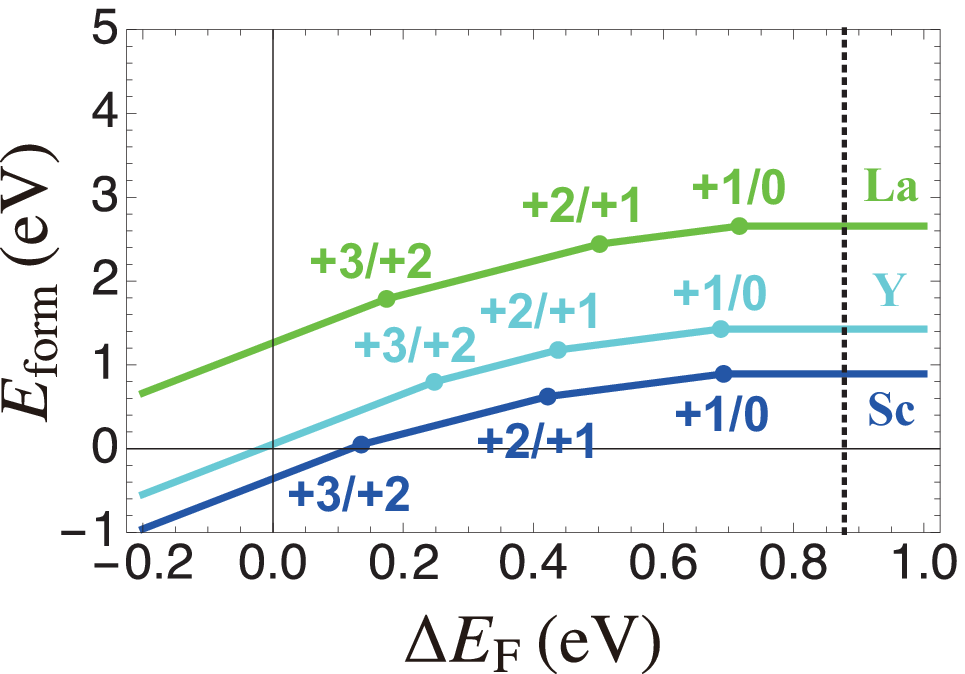} 
\caption{(Color online) 
Plots similar to FIG. \ref{fig:xAs} for CaZn$_2$P$_2$. 
The dotted line denotes the position of the conduction band minimum ($\Delta E_{\text{F}} = 0.878$ eV). 
} 
\label{fig:xP}
\end{center} 
\end{figure}

$E_{\text{form}}$ against the Fermi energy is displayed for CaZn$_2$As$_2$ and CaZn$_2$P$_2$ in FIG. \ref{fig:xAs} and FIG. \ref{fig:xP}, respectively. 
We calculate $E_{\text{form}}$ for several charged states ($q = 0, +1, +2, +3, +4$), 
and the least $E_{\text{form}}$ in these charged states is plotted against $\Delta E_{\text{F}}$ for each dopant case. 
The Fermi energy $\Delta E_{\text{F}} = E_{\text{F}} -E_{\text{V}}$ is measured from the valence band maximum $E_{\text{V}}$, 
so that  $\Delta E_{\text{F}} = 0$ is chosen to be the valence band maximum, 
while the conduction band minimum is located at $\Delta E_{\text{F}} = 0.408$ eV and $0.878$ eV for CaZn$_2$As$_2$ and CaZn$_2$P$_2$, respectively (see $E^{\text{GGA}}_{\text{g}}$ in Fig. \ref{fig:crystal_electronic}). 

When one focuses on the conduction band minimum to see the feasibility of the n-type (electron carrier) doping, 
it can be observed that the formation energies of $AE$ = Ca, Mg or $G3$ = Sc, Y are relatively small compared to the other dopants for both CaZn$_2X_2$ ($X$ = As, P), 
so that these dopants are relatively favorable for n-type doping within the interstitial doping of the dopants considered here. 
On the other hand, it appears that spontaneous doping of these dopants may be unlikely as far as these calculation results are concerned because these formation energies are all positive. 
One can also see that the relative magnitude of $E_{\text{form}}$ among the dopants is very similar between $X$ = As and P, i.e., Ca $<$ Mg $<$ Sr $<$ Ba and Sc $<$ Y $<$ La, 
and that each formation energy for CaZn$_2$As$_2$ is smaller than that for CaZn$_2$P$_2$ as a whole. 
Since $E_{\text{form}}$ at the conduction band bottom is always positive, it is not clear whether interstitial-site doping can actually result in n-type doping. 
Nonetheless, these  combinations might be possible candidates since $E_{\text{form}}$ for CaZn$_2$As$_2$ with $AE$ = Ca, Mg or $G3$ = Sc are relatively small. 

\subsection{$G3$ substitutional doping for Ca site} 
Next we consider the $G3$ (= Sc, Y, La) substitutional doping for the Ca site. 
As mentioned in the Introduction, element $A$ defects tend to arise in 122 Zintl phase compounds $AB_2X_2$,~\cite{Tamaki_2016,Ohno_2018,Pomrehn_2014,Peng_2018}  
so we assume that the introduction of dopant $D$ compensates for the Ca vacancy. 
Here the group 3 elements $G3$ (= Sc, Y, La) are chosen to be a dopant since they are expected to be cation $G3^{3+}$ substituting Ca$^{2+}$, and hence result in n-type doping.

\begin{figure}[t]
\begin{center}
\includegraphics[width=8.5cm,clip]{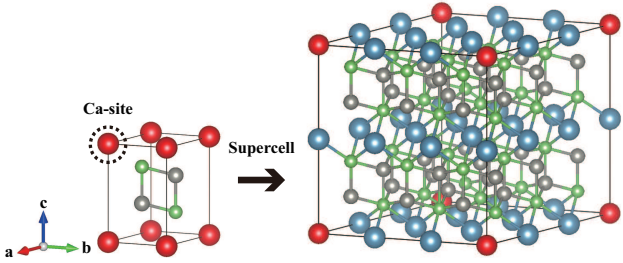} 
\caption{(Color online) 
(Left) Ca site of CaZn$_2X_2$ ($X$ = As, P). 
(Right) 3 $\times$ 3 $\times$ 2 supercell with a dopant (as an example). 
The red ball denotes the dopant, namely, $G3$ (= Sc, Y, La). 
} 
\label{fig:Supercell_Ca-site}
\end{center} 
\end{figure}

\begin{figure}[t]
\begin{center}
\includegraphics[width=8.5cm,clip]{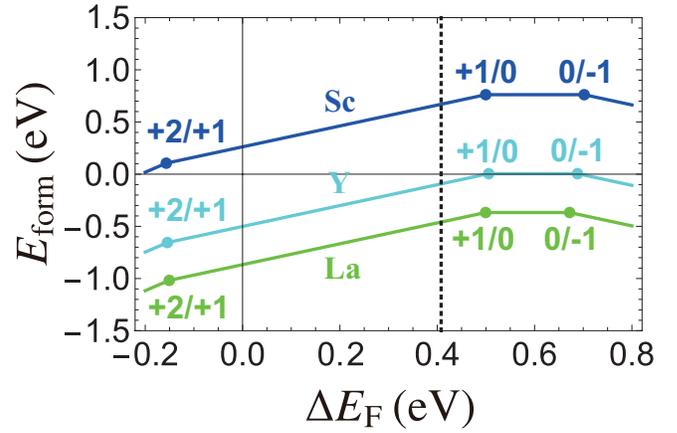} 
\caption{(Color online) 
The formation energy of the $G3$ substitutional doping for the Ca site $E_{\text{form}}$ against the Fermi energy $\Delta E_{\text{F}} = E_{\text{F}} -E_{\text{V}}$ for CaZn$_2$As$_2$. 
The group 3 elements $G3$ (= Sc, Y, La) are chosen as dopant $D$. 
The dotted line denotes the position of the conduction band minimum ($\Delta E_{\text{F}} = 0.408$ eV). 
} 
\label{fig:yAs}
\end{center} 
\end{figure}

\begin{figure}[t]
\begin{center}
\includegraphics[width=8.5cm,clip]{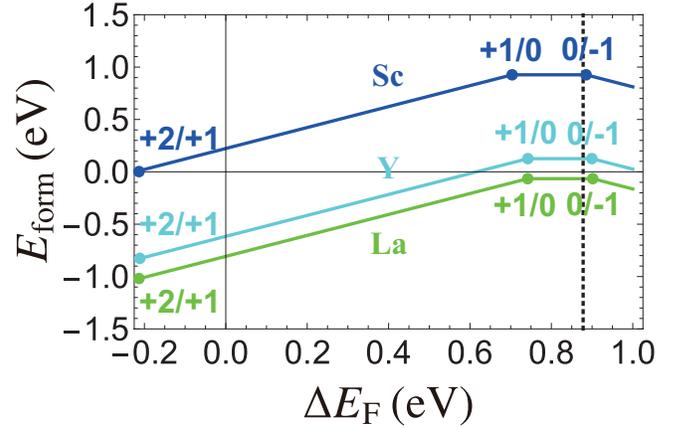} 
\caption{(Color online) 
Plots similar to FIG. \ref{fig:yAs} for CaZn$_2$P$_2$.
The dotted line denotes the position of the conduction band minimum ($\Delta E_{\text{F}} = 0.878$ eV). 
} 
\label{fig:yP}
\end{center} 
\end{figure}

The Ca site of CaZn$_2X_2$ and a supercell with a dopant (3 $\times$ 3 $\times$ 2 as an example) are displayed in FIG. \ref{fig:Supercell_Ca-site}. 
As in the case of the interstitial-site doping, 
the formation energy in this situation is now evaluated numerically from Eq. (\ref{eq:Eform}), as 
\begin{equation} 
\begin{split} 
E_{\text{form}} 
&=         \{ E[\text{(Ca$_{N-1}$$D$)Zn$_{2N}X_{2N}$}-qe^{-}] +E^{\text{corr}}_{1/L} \} +E[\text{Ca}]  \\ 
&\quad -E[\text{Ca$_N$Zn$_{2N}X_{2N}$}] - E[D]  \\ 
&\quad +q\left( E_{\text{V}} + \Delta E_{\text{F}} \right) , 
\end{split} 
\end{equation} 
where $E[\text{(Ca$_{N-1}$$D$)Zn$_{2N}X_{2N}$}-qe^{-}]$ represents the supercell of CaZn$_2X_2$, 
where Ca is substituted with the dopant $D$ in a charged state with charge $q$. 
The charged states $q= -1, 0, +1, +2$ are considered in the present case of the substitutional doping. 
Similarly to the case of the interstitial doping, $E[D]$ and $E[\text{Ca}]$ are determined as the energy of the bulk crystal per atom. 
$E_{\text{form}}$ against the Fermi energy is shown for CaZn$_2$As$_2$ and CaZn$_2$P$_2$ in FIG. \ref{fig:yAs} and FIG. \ref{fig:yP}, respectively. 

Again focusing on the conduction band minimum, one can observe that the formation energies of $G3$ = La, Y are relatively small compared to the other dopants. 
In particular, we notice that La can be substituted spontaneously for the Ca site since the formation energy is negative for CaZn$_2X_2$ ($X$ = As, P). 
As in the case of interstitial doping, the relative magnitude of $E_{\text{form}}$ among different dopants is the same between $X$ = As and P, 
and also $E_{\text{form}}$ tends to be smaller for $X$ = As than for $X$ = P. 

\section{Discussion} 
In the previous sections, 
we have considered interstitial-site doping of alkaline earth metals and group 3 elements $G3$, and $G3$ substitutional doping for the Ca site as a possible candidate for n-type doping. 
Both for CaZn$_2X_2$ ($X$ = As, P), the interstitial-site doping of $AE$ = Ca, Mg or $G3$ = Sc, Y, and $G3$ = La, Y substitutional doping are favorable in terms of energy stability, 
and also, each of these formation energies for CaZn$_2$As$_2$ is small compared to that of CaZn$_2$P$_2$. 
In particular, the formation energy for the La substitutional doping is the lowest among the considered cases both for CaZn$_2X_2$ ($X$ = As, P) 
and is negative, so that La is expected to be substituted for the Ca site and to provide electron carriers spontaneously. 

A relevant factor to the above mentioned tendency is considered to be the ionic radius of the dopants $AE$ and $G3$. 
For the interstitial-site doping, the formation energy becomes smaller with decreasing the ionic radius among $AE$ or $G3$ 
($r_{\text{Mg$^{2+}$}} < r_{\text{Ca$^{2+}$}} < r_{\text{Sr$^{2+}$}} < r_{\text{Ba$^{2+}$}}$ and $r_{\text{Sc$^{3+}$}} < r_{\text{Y$^{3+}$}} < r_{\text{La$^{3+}$}}$) 
except for Ca and Mg, where the formation energy is larger for the latter. 
For the $G3$ substitutional doping, the formation energy becomes smaller when the ionic radius of the dopant becomes closer to that of Ca.
In fact, the ionic radius of Ca ($r_{\text{Ca$^{2+}$}} \sim 100$ pm) is very close to that of La$^{3+}$ ($r_{\text{La$^{3+}$}} \sim 103$ pm),~\cite{Shannon_1969,Shannon_1976} 
and the formation energy of La substitutional doping for the Ca site is the lowest among the considered dopants. 

From the viewpoint of actual experiments, possibility of n-type doping of CaZn$_2$Sb$_2$ is also of great interest, 
but in the present study, we focused only on $X$ = As and P. 
As already mentioned, the reason for this is because the band gap closes for CaZn$_2$Sb$_2$, which is an artifact of GGA. 
Nonetheless, if we assume that the tendency found in the present study between $X$ = As and P also holds among $X$ = Sb, As and P, 
i.e., the formation energy behaves as CaZn$_2$Sb$_2$ $<$ CaZn$_2$As$_2$ $<$ CaZn$_2$P$_2$, 
CaZn$_2$Sb$_2$ would be more favorable for n-type doping compared to CaZn$_2$As$_2$ and CaZn$_2$As$_2$. 

In fact, since GGA generally underestimates the band gap, this would affect the formation energy, and even its sign could change. 
Therefore, even for CaZn$_2$As$_2$ and CaZn$_2$P$_2$, we cannot be completely certain about the possibility of spontaneous doping. 
Nevertheless, we believe that the overall systematic tendency among the dopants is grasped within the present study. 

\section{Summary} 
To summarize, we explored the possibility of n-type doping of CaAl$_2$Si$_2$-type Zintl phase compounds CaZn$_2X_2$ ($X$ = As, P) using first-principle calculations. 
Among the considered possibilities, 
the interstitial-site doping of $AE$ = Ca, Mg or $G3$ = Sc, Y, and the $G3$ = La, Y substitutional doping for the Ca have been found to have relatively small formation energies. 
In particular, the formation energy of the La substitutional doping is found to be negative for both CaZn$_2$As$_2$ and CaZn$_2$P$_2$. 
This suggests that La can substitute the Ca site spontaneously and hence provide electron carriers, as far as these calculation results are concerned. 
We have also found that the formation energies of the defects are smaller for CaZn$_2$As$_2$ than for CaZn$_2$P$_2$, 
which suggests that n-type doping is relatively easier for the former than for the latter.

\section*{Acknowledgments} 
We wish to thank H. Mori (RIKEN) and H. Usui (Shimane Univ.) for fruitful discussions. 
This work was supported by JST CREST (Grant No. JPMJCR20Q4) and JSPS KAKENHI Grant No. JP21K04866. 
Numerical calculations were done by the supercomputing systems 
at the Institute for Solid State Physics (ISSP) Supercomputer Center of the University of Tokyo 
and the Information Technology Center of the University of Tokyo.

\bibliographystyle{apsrev4-1} 

%

\end{document}